\documentclass[prd,preprint,tightenlines,floatfix,showpacs,preprintnumbers,nofootinbib,eqsecnum,superscriptaddress]{revtex4}

 \usepackage[dvips,final]{graphicx}
  \usepackage{amssymb}
   \usepackage{amsmath}
    \usepackage{amsfonts}
     \usepackage{epsfig}
      \usepackage{bm}

\usepackage[section]{placeins}

\usepackage{multirow}
\usepackage{ctable}
\usepackage{booktabs}
\usepackage{array}
\usepackage{tabularx}
\usepackage{xcolor}
\usepackage{pstricks}

\newcommand{\bb}{{\bf b}}
\def\lsim{\mathrel{\rlap{\lower4pt\hbox{\hskip1pt$\sim$}}
    \raise1pt\hbox{$<$}}}         
\def\gsim{\mathrel{\rlap{\lower4pt\hbox{\hskip1pt$\sim$}}
    \raise1pt\hbox{$>$}}}         

\begin{document}
\title{Electromagnetic excitation of nuclei and neutron evaporation \\ 
       in ultrarelativistic ultraperipheral heavy ion collisions}

\author{Mariola K{\l}usek-Gawenda}
\email{mariola.klusek@ifj.edu.pl} \affiliation{Institute of Nuclear
Physics PAN, PL-31-342 Cracow,
Poland}

\author{Micha{\l} Ciema{\l}a}
\email{michal.ciemala@ifj.edu.pl} \affiliation{Institute of Nuclear
Physics PAN, PL-31-342 Cracow,
Poland}

\author{Wolfgang Sch\"afer}
\email{wolfgang.schafer@ifj.edu.pl} \affiliation{Institute of Nuclear
Physics PAN, PL-31-342 Cracow,
Poland}

\author{Antoni Szczurek}
\email{antoni.szczurek@ifj.edu.pl} 
\affiliation{Institute of Nuclear
Physics PAN, PL-31-342 Cracow,
Poland}

\affiliation{
University of Rzesz\'ow, PL-35-959 Rzesz\'ow, Poland}

\date{\today}

\begin{abstract}
We present a new approach for calculating electromagnetic excitation
of nuclei as well as probabilities of emission and distributions of neutrons
from decays of excited nuclear systems for ultrarelativistic, ultraperipheral heavy ion collisions.
Excitation functions for $\gamma$ + Pb $\to$ Pb$^*$ are parametrized
using physics-motivated components: excitation of giant resonances,
quasi-deuteron absorption mechanism, excitation of nucleon resonances
as well as high-energy dissociation of protons and neutrons.
Neutron emission (up to 10 neutrons) from low-energy excitations
of $^{208}$Pb is modelled in terms of the Hauser-Feshbach formalism.
The probabilities of a given number of neutrons are calculated
as a function of excitation energy in a Monte-Carlo code GEMINI.
These functions are parametrized by smooth analytical functions. 
The results are compared to appropriate data 
for the $\gamma$ Pb $\to$ Pb$^*$ $\to$ k n reaction.
As an example the approach is used for calculating electromagnetic 
excitation in UPC processes. Both single photon and double photon excitations
are included and discussed. Topological cross sections with a given number
of neutrons in forward and backward directions are calculated in the calculation at the LHC energies. 
Excitation functions are presented. 
The results of calculation are compared with SPS and more recent ALICE experimental data and good agreement is achieved.
\end{abstract}


\maketitle

\section{Introduction}
\label{introduction}

The exclusive production of mesons, pairs of quark and antiquark, 
pairs of leptons or other Standard Model particles in ultraperipheral heavy ion collisions
(UPC) has been attracting much attention 
\cite{rewiev_articles, rewiev_articles_ours}. 
A measurement of such reactions at the high energies of present day colliders
(RHIC, LHC) often requires special triggers. Large charges of colliding ions 
lead to the production of huge fluxes of associated photons. These 
photons when scattered on the collision partner lead to its excitation. 
As will be discussed in the present paper low-energy excitations 
($E^* <$ 50~MeV) play especially important role.
The low-energy excited nuclear heavy systems, 
close to giant resonance region, 
decay predominantly via emission of neutrons. Because the energy
of the neutrons in the nucleus rest frame is rather small 
($\sim$ 10 - 20 MeV), in UPCs the neutrons are emitted in very small cones
around beam directions. Such neutrons can be registered by the so-called
zero-degree calorimeters (ZDC's) which are associated with many
high-energy detectors, such as e.g. STAR at RHIC \cite{ZDC_STAR} and 
ALICE at LHC \cite{ZDC_ALICE}.

In the present paper we wish to present our approach which
includes description of photoexcitation of nuclei and decay of excited
nuclei in the framework of Hauser-Feshbach theory. Results of our
calculations for $\gamma$ Pb $\to$ Pb$^*$ k n are confronted with
existing experimental data. Then topological cross sections
with a given number of neutrons in ion-ion collisions are calculated and 
compared to RHIC and LHC data. We discuss the role of single and double-photon 
excitations. We present simple parametrizations of the relevant impact parameter 
profiles which can be conveniently used for a multitude of central final states
produced in diffractive or $\gamma \gamma$ subprocesses, such as
vector mesons, leptons, pions, etc.

\section{Formalism}
\label{sec:formalism}

\subsection{Electromagnetic excitation in UPC}
\label{subsec:em_exc}

In this subsection we collect the classical probability calculus methods needed 
\cite{Llope:1990vp,Pshenichnov:2011zz,Baur:2003ar} to describe 
the electromagnetic excitation of ions in UPCs due to multiple photon exchanges.

\begin{figure}[!h]
\includegraphics[scale=0.4]{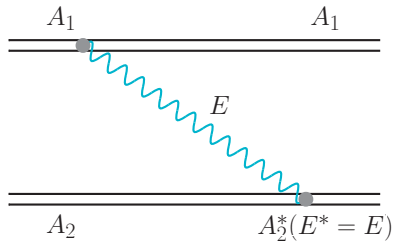}
   \caption{
\small Single excitation in UPCs.}
\label{fig:single_exc}
\end{figure}

\begin{figure}[!h]
\includegraphics[scale=0.4]{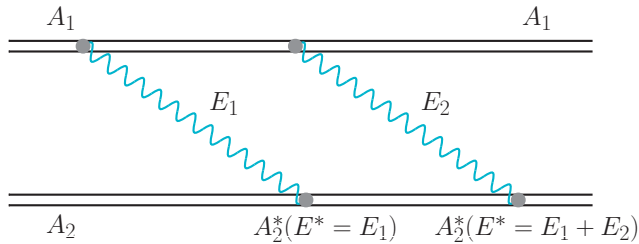}
   \caption{
\small Double excitation in UPCs.}
\label{fig:double_exc}
\end{figure}

From the usual Weizs\"acker-Williams flux of photons 
$N_A(E,\bb)$ and the total photoabsorption cross section
$\sigma_{\mathrm{tot}}(\gamma A; E)$ discussed in section \ref{subsec:photon_exc}, 
we introduce the mean number 
of photons absorbed by a nucleus $A_2$ in the collision with nucleus $A_1$:
\begin{eqnarray}
\bar n_{A_2} (\bb) \equiv \int_{E_{\mathrm{min}}}^\infty dE \,  
N_{A_1}(E,\bb)  \sigma_{\mathrm{tot}}(\gamma A_2; E) \, .
\end{eqnarray}
Here the upper limit in the integral is only formal: the photon flux (see any of the reviews \cite{rewiev_articles})
\begin{eqnarray}
N_A(E,\bb) = {1 \over E} \, { Z^2 \alpha_{\mathrm{em}} \over \pi^2} \, {1 \over \bb^2} \, \xi^2 K^2_1(\xi), \, \, \xi = {Eb \over \gamma_\mathrm{lab}} \, , 
\end{eqnarray}
implicitly contains a cut off in energy. 
Above $E$ is the photon energy in the rest frame of nucleus $A_2$, $Z$ is the nuclear charge and 
$K_1$ is a modified Bessel function. The boost to the rest frame of nucleus $A_2$ is given by
\begin{equation}
 \gamma_{\mathrm{lab}} = 2 \gamma^2_{\mathrm{cm}} -1 \, , \, \gamma_{\mathrm{cm}} = {\sqrt{s_{NN}} \over 2 m_N} \, .
\end{equation}
The lower limit
of integration $E_{\mathrm{min}}$ is the threshold for photoexcitation.
As in practice $\bar n_A(\bb) \ll 1$, for statistically independent 
absorption, we can state the probability of absorption of $n$ photons 
at impact parameter $\bb$ in the Poissonian form
\begin{eqnarray}
w_n(\bb) = {\Big(\bar n_A(\bb) \Big)^n \over n !} \, \exp[-\bar n_A(\bb)] \, .
\end{eqnarray}
We define the probability density for a single photon to excite nucleus $A_2$ 
in a collision at impact parameter $\bb$ of the $A_1-A_2$ collision as
\begin{eqnarray}
p^{(1)}_{A_2}(E,\bb) =   {N_{A_1}(E,\bb)  \sigma_{\mathrm{tot}}(\gamma A_2; E) \over \bar n_{A_2} (\bb)} \, ,
\end{eqnarray}
which is fulfilled, at each $\bb$
\begin{eqnarray}
  \int_{E_{\mathrm{min}}}^\infty dE \, p^{(1)}_{A_2}(E,\bb) = 1 \, .
\end{eqnarray}
Still under the assumption of statistical independence, $n$ photons 
will excite the nucleus with the probability density
\begin{eqnarray}
p^{(n)}_{A_2}(E,\bb)  = \int dE_1 dE_2 \dots dE_n \, \delta(E-\sum_{j=1}^n E_j) \,  p^{(1)}_{A_2}(E_1,\bb)  p^{(1)}_{A_2}(E_2,\bb) \dots  p^{(1)}_{A_2}(E_n,\bb) \, .
\end{eqnarray}
All the $n$-photon probability densities are properly normalized:
\begin{eqnarray}
  \int_{E_{\mathrm{min}}}^\infty dE \, p^{(n)}_{A_2}(E,\bb) = 1 \; .
\end{eqnarray}
Below we will explicitly calculate processes up to $n=2$ photon exchanges, see diagrams
in Figs. \ref{fig:single_exc},\ref{fig:double_exc}.

\begin{figure}[!h]
\includegraphics[scale=0.4]{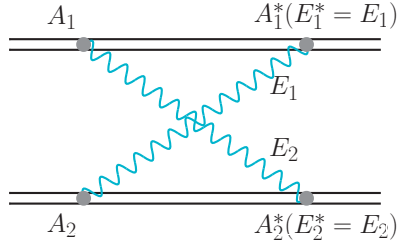}
   \caption{
\small Minimal mechanism for mutual excitation in UPCs.}
\label{fig:mutual_exc}
\end{figure}

Then, the probability for excitation of nucleus $A_2$ in the $n$-photon process is given by
\begin{eqnarray}
 w_n(\bb) p^{(n)}_{A_2}(E,\bb) \, .
\end{eqnarray}
We should sum over all numbers of photons
\begin{eqnarray}
{d P^{\mathrm{exc}}_{A_2} (\bb) \over dE} = \sum_n   w_n(\bb) p^{(n)}_{A_2}(E,\bb) \approx  \exp[-\bar n_{A_2}(\bb)] 
N_{A_1}(E,\bb)  \sigma_{\mathrm{tot}}(\gamma A_2; E)
 \, ,
\end{eqnarray}
where we indicated that we expect the single-photon absorption to
dominate. Notice that this may in practice depend on impact parameter.
The total probability for the nucleus to be excited is then 
\begin{eqnarray}
P^{\mathrm{exc}}_{A_2}(\bb) = \int dE  {d P^{\mathrm{exc}}_{A_2} (\bb) \over dE} = 1- \exp[-\bar n_{A_2}(\bb)] \approx \bar n_{A_2}(\bb) 
\exp[-\bar n_{A_2}(\bb)] \, .
\end{eqnarray}
The excitation cross section is then
\begin{eqnarray}
&&\sigma_{\mathrm{tot}}(A_1 A_2 \to A_1 A_2^*) = \int d^2\bb \, P_{\mathrm{surv}}(\bb) P^{\mathrm{exc}}_{A_2}(\bb) = \int d^2\bb  P_{\mathrm{surv}}(\bb)  \Big( 1- \exp[-\bar n_{A_2}(\bb)] \Big) \; .
\nonumber \\
\end{eqnarray}
Sometimes we are interested in the excitation cross section containing 
only excitations up to 
$E_{\mathrm{max}} \lsim 100 \, {\mathrm{MeV}}$, then we can
calculate the cross section from
\begin{eqnarray}
&&\sigma_{\mathrm{tot}}(A_1 A_2 \to A_1 A_2^*;E_{\mathrm{max}}) \approx
\int d^2\bb P_{\mathrm{surv}}(\bb) \exp[-\bar n_{A_2}(\bb)] \nonumber \\
&&\times \int_{E_{\mathrm{min}}}^{E_{\mathrm{max}}} dE \,  N_{A_1}(E,\bb)  \sigma_{\mathrm{tot}}(\gamma A_2; E) \, .
\end{eqnarray}
Here 
\begin{eqnarray}
P_{\mathrm{surv}}(\bb) \sim \theta(|\bb|-(R_{A_1}+R_{A_2})) \, , 
\end{eqnarray}
is the probability for the nuclei to survive the collision without 
additional strong interactions. As is apparent
\begin{eqnarray}
w_0(\bb) =  \exp[-\bar n_{A_2}(\bb)] \, ,
\end{eqnarray}
is the contribution to the survival probability from the electromagnetic
dissociation channels. The cross section for mutual electromagnetic 
dissociation is simply obtained from
\begin{eqnarray}
 &&\sigma_{\mathrm{tot}}(A_1 A_2 \to A_1^* A_2^*) = \int d^2\bb \, P_{\mathrm{surv}}(\bb) P^{\mathrm{exc}}_{A_2}(\bb) P^{\mathrm{exc}}_{A_1}(\bb) \, .
 \label{eq:em_diss}
\end{eqnarray}
In Fig. \ref{fig:double_exc} we show the situation when
two photons emitted by one of the colliding nuclei hit the
second nucleus.
Finally in Fig. \ref{fig:mutual_exc} we show as an example the case when
each of the nuclei emit a photon which excites then the collision partner. We shall call this case mutual excitations. 
The diagram shows a minimal mechanism needed to excite both nuclei simultaneously. Higher-order diagrams are possible too. 

\subsection{Dissociation into specific final states}
\label{subsec:dissociation}

\begin{figure}[!h]
\includegraphics[scale=0.4]{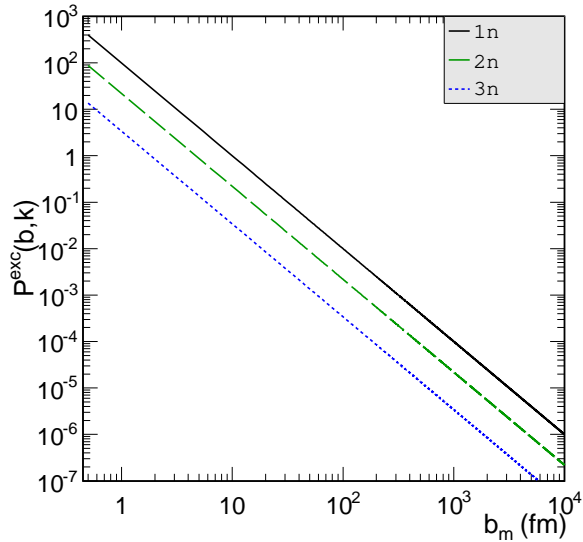}
   \caption{
\small Impact parameter profile for processes with evaporation of $k$ neutrons for $k=1,2,3$.}
\label{fig:P_b}
\end{figure}

In the present paper we are interested mainly in final states that contain a few neutrons, and want to study excitation cross sections as a 
function of neutron multiplicity (a type of ``topological cross sections''). Here the crucial input are 
the fractions $f(E,k)$ of a final state with $k$ neutrons coming from the decay of an excited nucleus at excitation energy $E$.
With their help, we can calculate the impact parameter profiles for processes with $k$ evaporated neutrons as:
\begin{eqnarray}
&& {d P^{\mathrm{exc}}_{A_2} (\bb,k) \over dE} = f(E,k) \cdot \sum_n   w_n(\bb) p^{(n)}_{A_2}(E,\bb) \approx f(E,k) \, p^{(1)}_{A_2}(E,\bb) 
\bar n_{A_2}(\bb)\exp[-\bar n_{A_2}(\bb)], \nonumber \\
\end{eqnarray}
and, correspondingly:
\begin{eqnarray}
&& P^{\mathrm{exc}}_{A_2}(\bb,k) = \int_{E_{\mathrm{min}}}^{E_{\mathrm{max}}} dE  {d P^{\mathrm{exc}}_{A_2} (\bb) \over dE} \, .
\end{eqnarray}
In Fig. \ref{fig:P_b} we plot these distributions as a function of impact parameter for $k=1,2,3$ at $\gamma_{\mathrm{cm}} = 1470$.
The cross section for $k$-neutron excitation is then
\begin{eqnarray}
\sigma(A_1 A_2 \to A_1 (k{\mathrm{N}},X)) = \int d^2\bb \, P_{\mathrm{surv}}(\bb) \, P^{\mathrm{exc}}_{A_2}(\bb,k)  \, .
\label{eq:single}
\end{eqnarray}
Of course we are confined to low-neutron multiplicities, as final states
of large number of neutron ($k$) can be produced by processes in the energy region 
$E > E_{\mathrm{max}}$ which we do not model so far.
Analogously, the mutual excitation cross sections with $m$ and $k$
neutrons in the debris of nucleus $A_1,A_2$, respectively, is
\begin{eqnarray}
\sigma(A_1 A_2 \to (m{\mathrm{N}},X) (k{\mathrm{N}},Y)) = \int d^2\bb \, P_{\mathrm{surv}}(\bb) \, P^{\mathrm{exc}}_{A_1}(\bb,m) \,  P^{\mathrm{exc}}_{A_2}(\bb,k) \, .
\label{eq:mutual}
\end{eqnarray}

\section{Photon-induced excitation of nuclei and neutron evaporation}
\label{subsec:photon_exc}

To evaluate the photoabsorption probabilities, we need a parametrization
of the total phtotabsorption cross section over a broad range of
energies. Here we are not interested in a microscopic
modelling of the different mechanisms play an important role at different
energies, but rather in a fit of empirical data.

At the lowest energies of relevance, photoabsorption 
is dominated by giant resonances.

The energy dependence of the cross section
for the giant dipole resonance (GDR) component ($\sigma_{\mathrm{GDR}}$) 
is parametrized following Ref.~\cite{Speth,Plujko}:
\begin{eqnarray}
 \sigma_{\mathrm{GDR}} = \frac{2}{\pi} \sigma_{TRK}  \frac{E^2 \Gamma_r}{\left( E^2-E_r^2 \right)^2+\left( E \Gamma_r \right)^2} S_r \; .
\end{eqnarray}
The parameters 
\begin{eqnarray}
\sigma_{TRK} = 60 \frac{NZ}{A} \mbox{ mb MeV}, 
E_r = 13.373 \mbox{ MeV}, 
\Gamma_r = 3.938 \mbox{ MeV},
S_r = 1.33716.
\end{eqnarray}
are taken from Ref.~\cite{GDRparameters}.

At somewhat larger energies a so-called quasi-deuteron contribution plays
important role and following \cite{Chadwick} is parametrized as
\begin{eqnarray}
\sigma_{\mathrm{QD}} = 6.5 \frac{NZ}{A} \sigma_d f(E) \; ,
\end{eqnarray}
where
\begin{equation}
\sigma_d = 61.2 \frac{\left( E-2.224 \right)^{3/2}}{E^3} \mbox{ mb} \; ,
\end{equation} 
\begin{eqnarray}
f(E<20 MeV ) &=& \exp \left( -73.3/E \right), \nonumber \\
f(20<E<140 MeV) &=& 8.3714 \times 10^{-2} \nonumber \\
                &-& 9.8343 \times 10^{-3}E + 4.1222 \times 10^{-4} E^2 \nonumber \\
                &-& 3.4762 \times 10^{-6} E^3 + 9.3537 \times 10^{-9} E^4\; , \nonumber \\
f(E>140 MeV) &=& \exp \left( -24.2/E \right) \; .
\end{eqnarray}
Above a photon energy $E_{\gamma} >$ 100 MeV the nucleon resonances are taken into account. The $\Delta$ resonance being the dominant feature of the 
excitation spectrum. 
We parametrize this region of the photoabsorption cross section with the help of 
a Gaussian function
\begin{equation}
\sigma_{Gauss} = \frac{C_G}{\sigma_G \sqrt{2\pi}} 
				\exp\left( \frac{-\left( E-\mu_G \right)^2}{2\sigma_G^2} \right) \; ,
\end{equation} 
where $C_G = 23 \, \mathrm{barn \,  MeV}$, $\sigma_G$ = 110 MeV, $\mu_G$ = 350~MeV.

Above $E_{\gamma} >$ 0.5 GeV the resonant contributions disappear
and the continuum related to break-up of nucleons starts to be important.
The corresponding total cross section (forward amplitude of photon elastic
scattering) is usually parametrized by exchange of pomeron at very large
energies and subleading reggeons at intermediate energies.
In our simple parametrization the pomeron exchange contribution
is parametrized as a constant and slightly phenomenological
function is used to represent the reggeon exchange contribution.  

\begin{figure}[!h]
\includegraphics[scale=0.4]{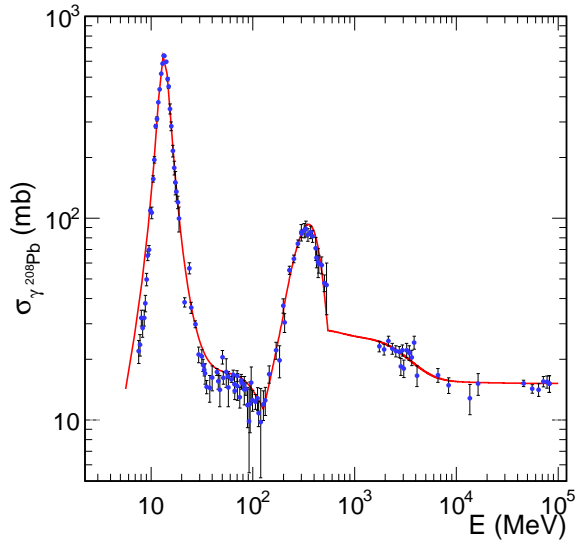}
   \caption{
\small Photoabsorption cross section for the $\gamma^{208}$Pb $\to$ $^{208}$Pb reaction}
\label{fig:photoabsorption}
\end{figure}

For the highest energy part ($E_{\gamma} >$ 8 GeV) we use a simple form
given in \cite{Vidovic} (below $\omega_0 = 80 \, \mathrm{GeV}$)
\begin{eqnarray} 
\sigma_{\gamma A}  = \Big( 15.2+0.06 \ln ^2 \left( \frac{E}{\omega_0} \right) \Big) \, \mathrm{mb},
\end{eqnarray}
This multicomponent parametrization is compared to the 
experimental data for photoabsorption on lead \cite{photoabsorption_data}
in Fig. \ref{fig:photoabsorption}.
The quality of the description of the data is fully sufficient for 
our purpose.

\subsection{Decays of excited nuclear systems}
\label{subsec:decays}

The calculation of probability of evaporated neutron multiplicity as a
function of $^{208}$Pb excitation energy was performed with 
the help of a Monte Carlo code GEMINI++~\cite{gemini}.  
In this code the evaporation process is described by Hauser-Feshbach 
formalism~\cite{refhaus}, in
which the decay width for evaporation of a particle $i$ from 
the compound nucleus with excitation energy $E^*$ and spin $S_{CN}$ is:
\begin{align}
\Gamma_i = \frac{1}{2\pi \rho (E^*, S_{CN})} \int d \epsilon \sum_{S_d=0}^{\infty} 
\sum_{J=|S_{CN}-S_d|}^{S_{CN} -S_d} \sum_{\ell=|J-S_i|}^{J+S_i} 
T_\ell (\epsilon) \rho(E^*-B_i - \epsilon, S_d),
\end{align}
where $S_d$ is the spin of the daughter nucleus, $S_i$, $J$ and $\ell$
are spin, total and angular momentum of the evaporated particle,
$\epsilon$, $B_i$ are kinetic and separation energies, $T_\ell$ is its 
transmission coefficient, $\rho$ and $\rho_{CN}$ are level densities of 
the daughter and compound nucleus, which can be calculated from the formula:
\begin{align}
\rho(E^*,S) \propto (2S +1) \exp\left(2\sqrt{a(U,S)U}\right),
\end{align}
where $U = E^* - E_{rot}(S) - \delta P$ is thermal excitation energy calculated by taking into account pairing corrections to the empirical mass formula ($\delta P$) and rotational energy $E_{rot}(S)$. 
In calculations the separation energies $B_{i}$, nuclear masses, shell and pairing corrections were used according to Ref.~\cite{moller}.
Level density parameter $a(U,S)$ was calculated as:
\begin{align}
a(U,S) = \tilde{a}(U) \left ( 1 - h(U/\eta +S/S_\eta) \frac{\delta W}{U}   \right),
\end{align}
where $\delta W $ is the shell correction to the liquid-drop mass and $\tilde{a}$ is smoothed level-density parameter, 
the function specifying the rate of fadeout is $h(x) = \tanh{x}$, the fadeout parameter $\eta $ 
was equal 18.52~MeV  and the parameter $S_\eta$ was set to 50~$\hbar$.
 
The smoothed level density parametrization depends on the nuclei excitation energy as:
\begin{align}
\tilde{a}(U) =\frac{A}{k_{\infty} - (k_\infty -k_0)\exp\left(-\frac{\kappa}{k_\infty - k_0}\frac{U}{A} \right)},
\end{align}
where $k_0 = 7.3$, $k_{\infty}=12$ and   $\kappa = 0.00517\exp(0.0345 A)$~\cite{gemini}.

\begin{figure}[!h]
\includegraphics[scale=0.4]{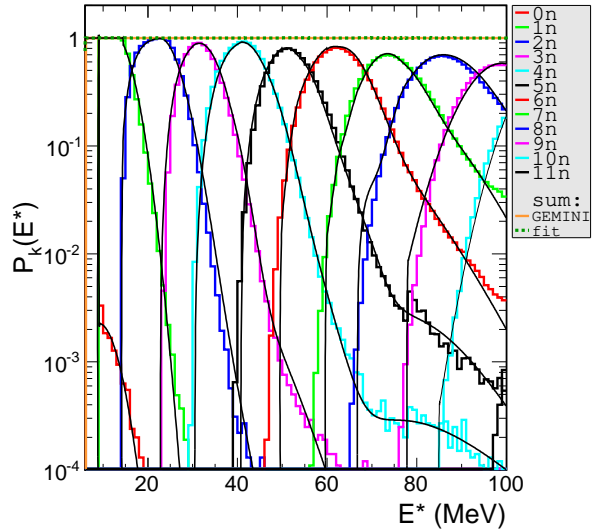}
   \caption{
\small Probability of neutron multiplicity as a function of excitation energy~($E^*$) of $^{208}$Pb nuclei.}
\label{fig:P_k}
\end{figure}

We assume that excited nucleus is 
formed with angular momentum equal to 0 (which we believe is a good approximation for photoproduction) and full energy is used for excitation.
The calculation is done with energy step of 1~MeV. 
For each excitation energy $10^6$ events (decays) were generated.
Finally neutron emissions probabilities were obtained from the MC
sample for each excitation energy (see histogram in Fig.~\ref{fig:P_k}).

The fractions of events with a $k$-neutron final state at excitation energy $E^*$ can
be well fitted by a sum of the following purely empirical functions:
\begin{equation}
 f(E^*,k) = f^{exp} (E^*,k) + f^{Gauss} (E^*,k) \; .
\end{equation}
\begin{equation}
 f^{exp} (E^*,k) = C_e \left( E^*-\mu_e \right)^2 \exp \left( \frac{-\left( E^*-\mu_e \right)}{\sigma_e} \right) \; ,
\end{equation}
\begin{equation}
 f^{Gauss} (E^*,k) = \frac{C_G}{\sigma_G \sqrt{2\pi}} \exp\left( \frac{-\left( E^*-\mu_G \right)^2}{2\sigma_G^2} \right) \; .
\end{equation}
The parameters of the phenomenological functions found in the fit are collected in Table \ref{tab:parameters_neutrons}
and can be used in any calculations.

To ensure that probabilities always add up to unity, in practice we impose
\begin{eqnarray}
f(E^*,2) &&= 1 - f(E^*,1) \, \, \mathrm{for} \, E^* < 22 \, \mathrm{MeV},
\nonumber \\
f(E^*,3) &&= 1 - f(E^*,1) - f(E^*,2) \, \, \mathrm{for} \, E^* < 30 \, \mathrm{MeV} \, ,
\end{eqnarray} 
and similarly for higher $k$.

\begin{table}
\caption{Parameters of fit functions for probability of neutron multiplicity $P_k(E^*)$.}
\label{tab:parameters_neutrons}
\begin{center}
\begin{tabular}{|c||c|c|c||c|c|c|}
\hline
 Number of neutrons	& $C_e$	& $\mu_e$	& $\sigma_e$	& $C_G$	& $\mu_G$	& $\sigma_G$	\\ \hline \hline
	0		&   0	& -		& -		& 0.02	& 9		& 3.5		\\ \hline
        1		&   0	& -		& -		& 9.9	& 12		& 3.5		\\ \hline
        2		&  0.08	& 21.6		& 1.7		& 10.3	& 21.5		& 4.2		\\ \hline
        3		&  0.0015 & 38		& 1.2		& 9.4	& 31.5		& 4.2		\\ 
			& 0.0005 & 40		& 2.6		&	&		&		\\ \hline
        4		&  0.0012 & 45		& 2.3		& 11	& 41		& 4.8		\\ 
			& 0.000008 & 62		& 8		&	&		&		\\ \hline
        5		&  0.03	& 53		& 2.6		& 8.7	& 51		& 4.3		\\ 
			& 0.00015 & 72		& 4.9		&	&		&		\\ \hline
	6		& 0.023	& 62.2		& 2.8		& 11.5	& 61.8		& 5.5		\\ 
			& 0.003 & 68		& 4.35		&	&		&		\\ \hline
	7		&  0.015 & 76		& 4		& 9.5	& 73.5		& 5.3		\\ \hline
	8		&  0	& -		& 	-	& 8.8	& 84		& 5.3		\\ \hline
	9		&  0	& -		& 	-	& 10.5	& 99		& 7.2		\\ \hline
	10		&  0	& -		& 	-	& 10.1	& 110		& 6.5		\\ \hline
\end{tabular}
\end{center}
\end{table}

\subsection{Excitation functions for the $\gamma$ Pb $\to$ Pb$^*$ $\to$ k n reaction}
\label{subsec:exc_functions}

\begin{figure}[!h]
\includegraphics[scale=0.4]{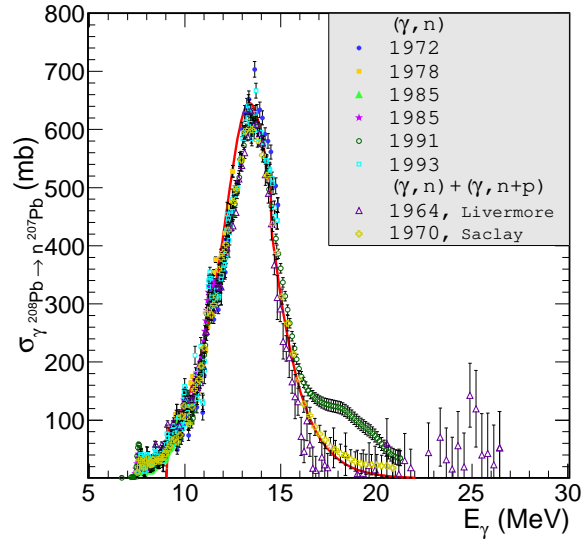}
   \caption{
\small Excitation energy for the $\gamma^{208}$Pb $\to$ n$^{207}$Pb
reaction. Experimental data are from Refs.
~\cite{1n:1972, 1n:1978, 1n:1985_1, 1n:1985_2, 1n:1991, 1n:1993, Livermore_1964, Saclay_1970}.
}
\label{fig:1n}
\end{figure}

\begin{figure}[!h]
\includegraphics[scale=0.4]{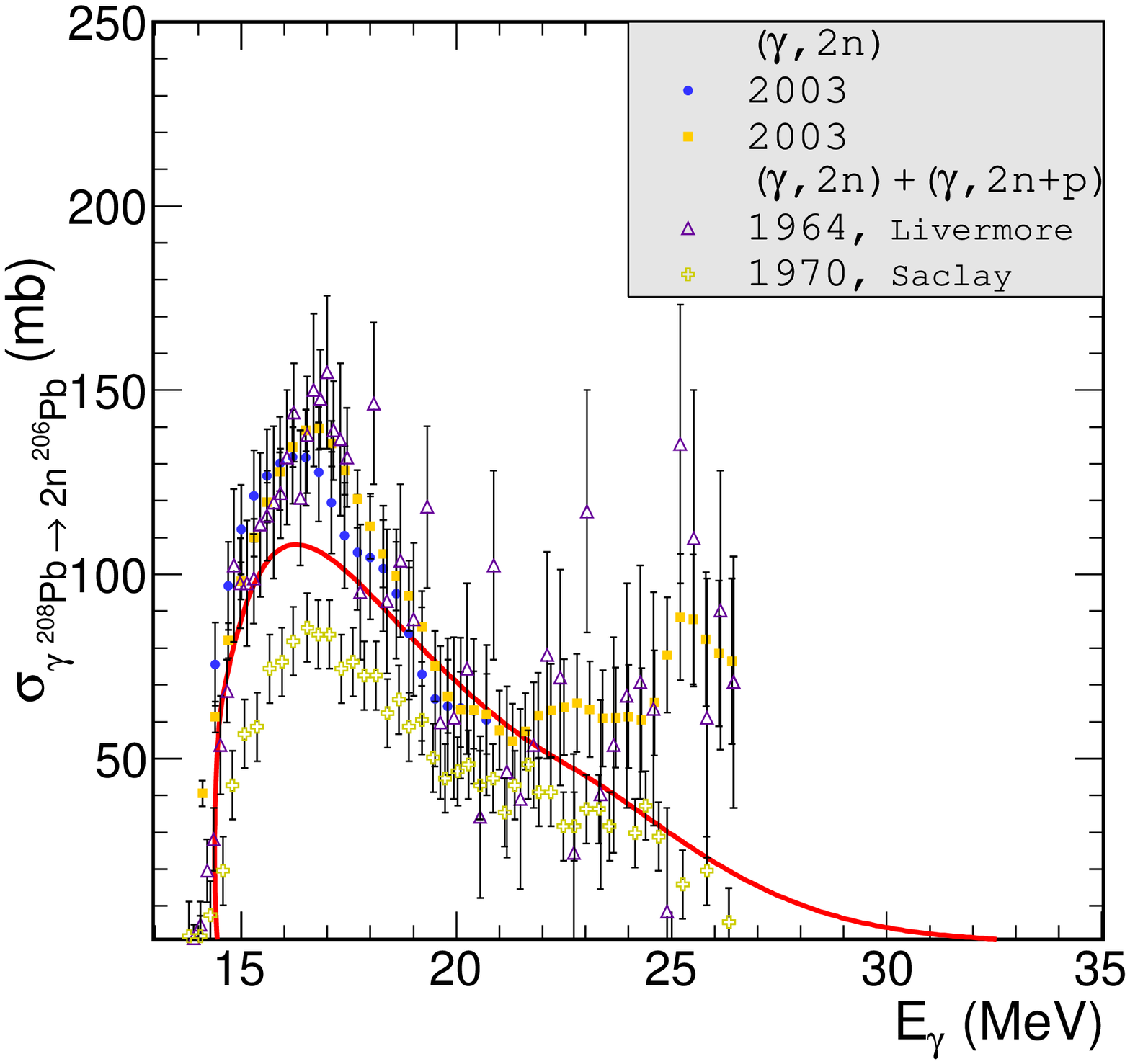}
   \caption{
\small Excitation energy for the $\gamma^{208}$Pb $\to$ 2n$^{206}$Pb
reaction. Experimental data are from Refs.
~\cite{2n:2003, Livermore_1964, Saclay_1970}.
}
\label{fig:2n}
\end{figure}

\begin{figure}[!h]
\includegraphics[scale=0.4]{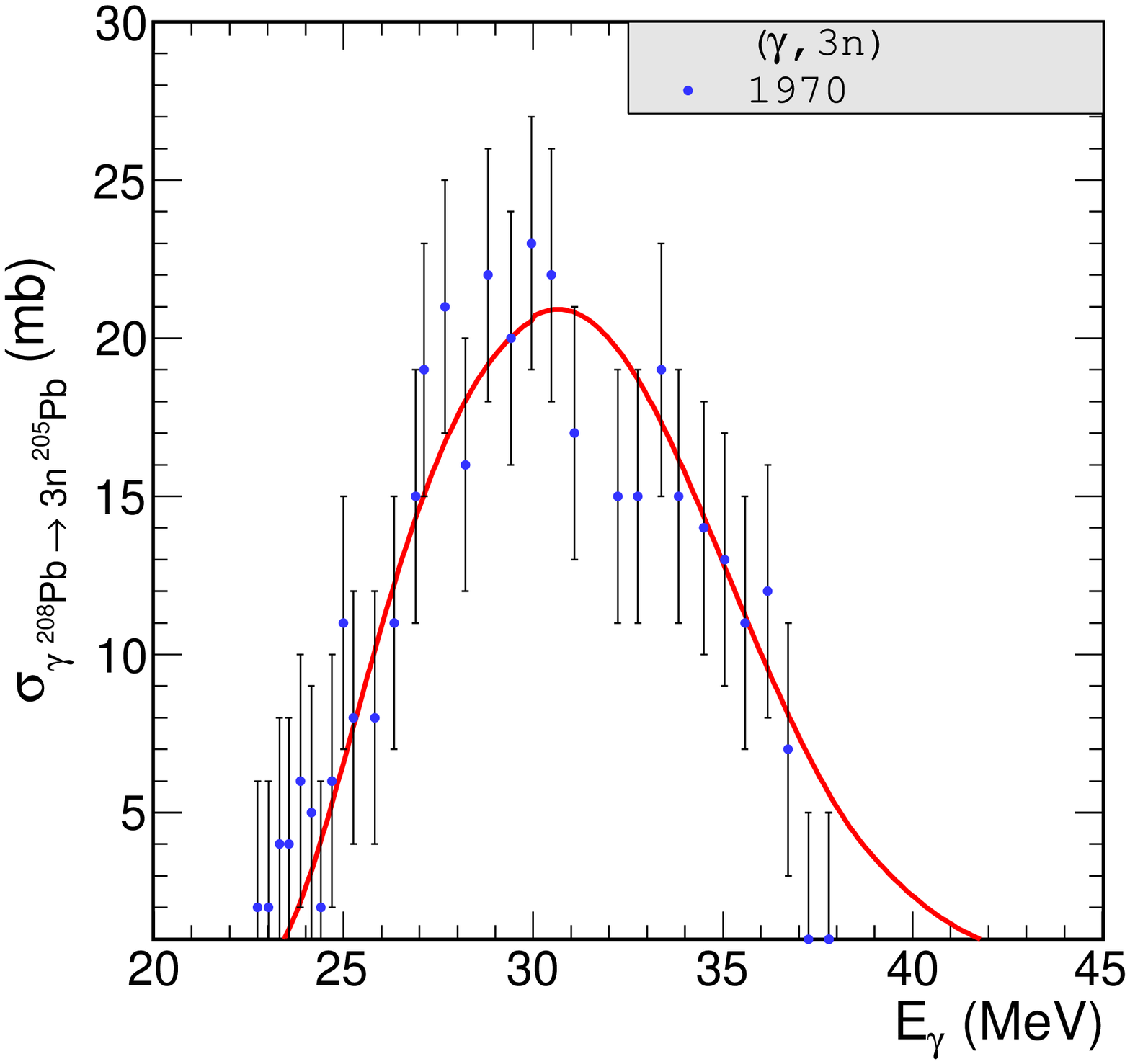}
   \caption{
\small Excitation function for the $\gamma^{208}$Pb $\to$ 3n$^{205}$Pb
reaction. Experimental data are from Ref.~\cite{Saclay_1970}.
}
\label{fig:3n}
\end{figure}

Using photoabsorption cross section shown in Fig.~\ref{fig:photoabsorption}
and probability to emit a fixed number of neutrons (k)
obtained as described in section \ref{subsec:photon_exc} we can calculate
photon-induced excitation function with a given number of associated
neutrons. The results are shown in Fig.~\ref{fig:1n}, \ref{fig:2n}, \ref{fig:3n}.

%
%

Quite a good agreement with the world data is obtained.
This is quite surprising given that our calculation implicitly assumes
equilibration of the nuclear system (Hauser-Feshbach formalism) 
formed after absorption of the photon. If we assumed that part of 
the energy of the photon would escape before equilibration of 
the nuclear system (due to pre-equilibrium processes) the agreement with
the data would be much worse. Having proven usefulness of our approach
we can proceed to the excitation of nuclei in UPCs and
related production of neutrons from the excited nuclear systems. 
In the following we shall present the results of the formalism discussed above.

\section{Results for electromagnetic excitations in UPCs}
\label{sec:EM_exc_UPC}

Now we shall present our results for electromagnetic excitation of
nuclei in UPCs.

In Table \ref{table:single_excitation} we have collected
cross section in barns for single-nucleus single-photon excitation 
for different ranges of excitation energy
and different collision energies represented by different
$\gamma_{\mathrm{cm}}$ adequate for RHIC and LHC. In both cases the calculation was done for lead nuclei.
Our results are compared with an earlier calculation by Vidovi\'c et al. \cite{Vidovic}. 
Very good agreement can be observed.
%
\begin{table}[h!]
\caption{Cross section in barns for single-nucleus, single-photon
  excitation for different ranges of excitation energy 
  for $^{208}$Pb + $^{208}$Pb collisions.}
	\label{table:single_excitation}
\begin{tabular}{|c||c|c|c|c}
\hline
			& (6-40) MeV	&	(40-2000) MeV	&	(2-80) GeV \\  \hline
\multicolumn{4}{|c|}{$\gamma_{\mathrm{cm}}$ = 100}\\  \hline
M.Vidovi\'c et al. \cite{Vidovic}	& 77.6		&	25.7		&	5.6		\\
Our results		& 80.16		&	25.6		&       5.6	\\ \hline

\multicolumn{4}{|c|}{$\gamma_{\mathrm{cm}}$ = 3100}\\ \hline
M.Vidovi\'c et al. \cite{Vidovic}	& 133.6		&	53.7		&	18.7		\\
Our results		& 133.8	        &	54.6		&       18.7	\\ \hline
\end{tabular}
\end{table}
As already discussed, we are very interested also in calculating
associated neutron multiplicities. In Table 
\ref{table:single_excitations_neutrons} we have collected appropriate
cross sections for one-photon single nucleus excitation in barns for different neutron multiplicities
k = 0, 1, 2, 3. We compare results with and without the exponential
factor in Eq.(\ref{eq:single}). As can be seen from Table \ref{table:single_excitations_neutrons} 
the exponential factor plays here only a minor role in practical calculations.
We have also collected experimental data of the ALICE collaboration \cite{ALICE2012}.
We observe some disagreement especially for 3 neutrons.
For the ratio $2n/1n$  we obtain $22 \%$, in good agreement with the ALICE result of 
$22.5 \pm  0.5 \, \mathrm{stat} \, \pm  0.9 \, \mathrm{syst} \, \%$.
%
\begin{table}[h!]
\caption{Cross section in barns for a given multiplicity of neutrons
in single-nucleus, single-photon excitation in $^{208}$Pb + $^{208}$Pb collisions at $\sqrt{s_{NN}}$ = 2.76 TeV.}
\label{table:single_excitations_neutrons}
\begin{center}
\begin{tabular}{|c||r|r|r|}
\hline
\multicolumn{4}{|c|}{Single excitations [b]}	\\ \hline \hline
		&	  Our results	&	$\exp(-\bar n)$ 	& 	ALICE (Ref.~\cite{ALICE2012})	\\ \hline
0 neutrons	&	  6.403		&	without		&			\\
		&	  6.394		&	with		&			\\ \hline
1 neutron	&	 84.301		&	without		&	~93.0		\\
		&	 82.888		&	with		&			\\ \hline
2 neutrons	&	 18.608		&	without		&	~21.0		\\
		&	 18.532		&	with		&			\\ \hline
3 neutrons	&	  2.858		&	without		&	~~6.5		\\ 
		&	  2.856		&	with		&			\\ \hline \hline
sum		&	112.170		&	without		&			\\ 
		&	110.670		&	with		&			\\ \hline \hline
total		&	
					&			&	187.4		\\ \hline
\end{tabular}
\end{center}
\end{table}

Two-photon exchanges may also lead to simultaneous excitation 
of both nuclei (see Fig.~\ref{fig:mutual_exc}). 
In Table \ref{table:mutual_excitations} we have collected
topological cross sections with a given number of neutrons emitted from
first ($k_1$) and second ($k_2$) nucleus. As previously we show results 
of the calculation with and without the extra exponential factor, which seems
here more important than for single-nucleus single-photon excitations.
This indicates an importance of smaller impact parameters in the two-photon
processes.
Our results could be compared to those in
Ref.~\cite{Pshenichnov:2011zz}. Compared to Ref.~\cite{Pshenichnov:2011zz} 
our cross section for neutron multiplicities $k_1$ = 1 and $k_2$ = 1 
are larger by about 25 \%.
Other numbers seem to be in much better agreement.
The differences quantify the  uncertainties of theoretical calculations.

\begin{table}[h!]
\caption{Cross section in barns for mutual excitation with
a given number of neutrons emitted from both nuclei in $^{208}$Pb + $^{208}$Pb collisions at $\sqrt{s_{NN}}$ = 2.76 TeV.}
\label{table:mutual_excitations}
\begin{center}
\begin{tabular}{|c||c|c|c|c|c|}
\hline
			\multicolumn{6}{|c|}{Mutual excitations [b]}	 \\  \hline
		&	0 neutrons	&	1 neutron	&	2 neutrons	&	3 neutrons	&	$\exp(-\bar n)$ 	\\ \hline \hline
0 neutrons	&	0.00917		&	0.12093		&	0.02675		&	0.00414		&	without		\\
		&	0.00883		&	0.09317		&	0.02483		&	0.00402		&	with		\\ \hline
1 neutron	&	0.12093		&	1.59450		&	0.35275		&	0.05448		&	without		\\
		&	0.09317		&	1.00124		&	0.26286		&	0.04238		&	with		\\ \hline
2 neutrons	&	0.02675		&	0.35275		&	0.07803		&	0.01205		&	without		\\
		&	0.02483		&	0.26286		&	0.06989		&	0.01130		&	with		\\ \hline
3 neutrons	&	0.00413		&	0.05448		&	0.01205		&	0.00186		&	without		\\
		&	0.00402		&	0.04238		&	0.01130		&	0.00183		&	with		\\ \hline \hline
sum		&	0.16098		&	2.12266		&	0.46958		&	0.07252		&	without		\\
		&	0.19285		&	1.39965		&	0.36888		&	0.05953		&	with		\\ \hline 
		&	\multicolumn{4}{|c|}{2.82574}								&	without		\\
		&	\multicolumn{4}{|c|}{2.02091}								&	with		\\ \hline \hline
\end{tabular}
\end{center}
\end{table}
%
\begin{table}
\begin{center}
\begin{tabular}{|c|}
\hline
Mutual excitations [b] \\
ALICE (Ref. \cite{ALICE2012}) \\  \hline
5.7 	\\ \hline 
\end{tabular}
\end{center}
\end{table}
%
In Table \ref{table:double_photon} we have collected contributions
to the double-photon excitation cross section when each of the two photons is in different
energy  intervals (the same integrals as defined previously in Table \ref{table:single_excitation}).
The nine different regions in (E$_1$ $\times$ E$_2$) space give comparable contribution to the total
cross section for double-excitation of a single nucleus.
\begin{table}[h!]
\caption{Cross section in barns for double-photon excitation
of one of nuclei in $^{208}$Pb + $^{208}$Pb 
collisions at $\sqrt{s_{NN}}$ = 2.76 TeV.}
\label{table:double_photon}
\begin{center}
\begin{tabular}{|l||r|r||r|r|}
\hline
				& $E_1$ = (6 - 40) MeV	& $E_1$ = (40 - 2 000) MeV	& $E_1$ = (2 - 80) GeV	& sum 	\\ \hline
$E_2$ = (6 - 40) MeV		&	2.893  		&	1.438		&	0.723		& 5.054	\\
$E_2$ = (40 - 2 000) MeV	&	1.438		&	0.716  		&	0.357		& 2.511	\\
$E_2$ = (2 - 80) GeV		&	0.723 		&	0.357		&	0.178		& 1.258	\\ \hline \hline
			&	5.054	&	2.511		&	1.258		& 8.823	\\ \hline \hline
\end{tabular}
\end{center}
\end{table}
It is particularly interesting how nuclei are excited in UPCs.

\begin{figure}[!h]
\includegraphics[scale=0.4]{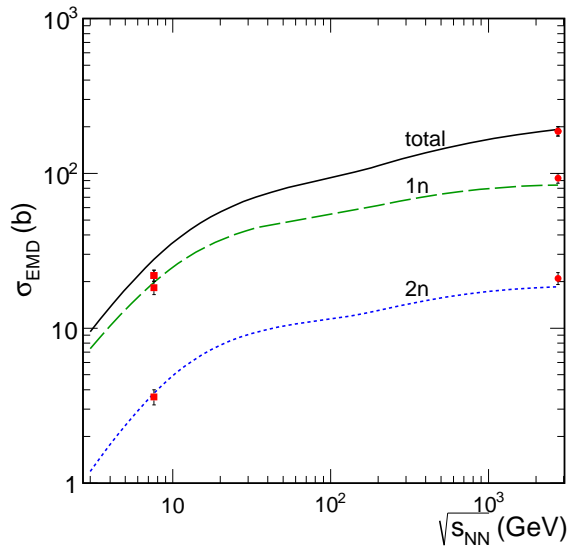}
   \caption{
\small Single EMD cross sections as the function of $\sqrt{s_{NN}}$. Data are from
SPS \cite{Golubeva:2005nq} and LHC (ALICE) \cite{ALICE2012}.
}
\label{fig:EMD_sNN}
\end{figure}

In Fig. \ref{fig:EMD_sNN} we show the total cross section for  electromagnetic excitation as a function of $\sqrt{s_{NN}}$ as well as the partial cross sections 
into one and two-neutron final states. 
It should be noted that we concentrate only on the neutrons
evaporated from the electromagnetically excited nuclei. We do not account
for neutrons from other hadronic processes, like the intranuclear cascading
(see for example \cite{Pshenichnov:2001qd, Pshenichnov:2011zz}).
We also neglect the mutual excitation of nuclei by strong interactions. 

\begin{figure}[!h]
\includegraphics[scale=0.4]{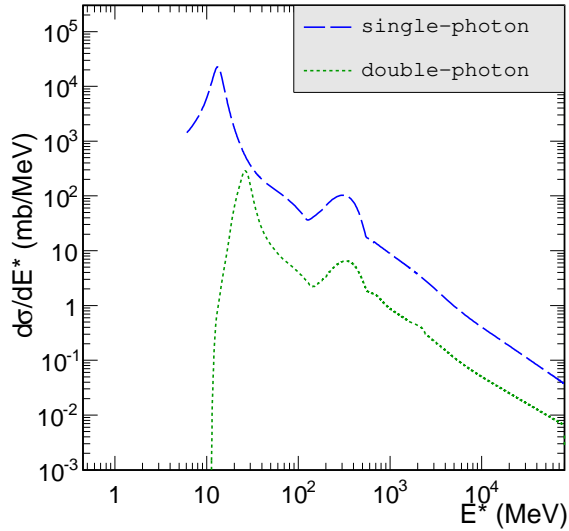}
 \caption{
\small Excitation function $\frac{d\sigma}{dE^*}$
for electromagnetic excitation of single nucleus in single-photon
(dashed) and double-photon (dotted) exchange
in UPCs at $\sqrt{s_{NN}}$ = 2.76 TeV.  
}
\label{fig:dsig_dE_UPC}
\end{figure}

In Fig. \ref{fig:dsig_dE_UPC} we show our result for
$^{208}$Pb + $^{208}$Pb $\to$ $^{208}$Pb$^*$ + $^{208}$Pb reaction at the LHC
energy $\sqrt{s_{NN}}$ = 2.76~TeV. For technical reasons different
regions (low (6-40 MeV), intermediate (40-2000 MeV), high ($>$2 GeV)) 
of excitation energy were calculated separately. 
The dashed line represents contribution of single-photon excitation (diagram in Fig. \ref{fig:single_exc}) and 
the dotted line double-photon excitation (diagram in Fig. \ref{fig:double_exc}).
Even at the very high collision energy ($\sqrt{s_{NN}}$ = 2.76 TeV) the
low-energy excitations are still essential. Please note, however, 
the logarithmic scale for the excitation energy axis,
which emphasises the low-energy excitation.
The double-photon excitation contribution is much smaller than
the single-photon one. In addition, the highest peak appears at the
excitation energy twice larger than for single-photon excitation, which
corresponds to excitation of giant dipole resonance excited on top of an already
excited one.
Such processes were already discussed in the literature (\cite{Pshenichnov:2011zz}
and the references therein).

\section{Conclusions}

In this paper we have presented a new approach for calculating
excitation of lead nuclei in photoabsorption reactions as well as
in ultraperipheral ultrarelativistic heavy ion collisions.
The photoabsorption cross section on lead nuclei is fitted using physics
motivated multicomponent parametrization. The giant resonances,
quasi-deuteron, excitation of nucleon resonances and break up of
nucleon mechanism are included in the fit to the world data.
Good quality fit is obtained.

The neutron emission from the excited nuclear system is calculated 
within the Hauser-Feshbach formalism.
Within our approach we get a very good description of excitation
functions for $\gamma$ + $^{208}$Pb with a fixed number of neutrons. 
The excitation function is used next to calculate cross sections 
in peripheral UPCs.
Both single and double-photon excitation processes are included and discussed.
We have calculated corresponding excitation functions for both single 
(one-nucleus) and mutual (both nuclei) excitations. 

We have obtained a good agreement of the calculated total cross section for electromagnetic excitation as well as cross section for one-neutron and two-neutron emission with recent experimental data of ALICE collaboration.

The formalism presented here may be easily applied to other exclusive 
ultrarelativistic heavy ion processes such as:
$A A \to A A J/\Psi$, $A A \to A A \rho^0$, $A A \to A A e^+ e^-$,
$A A \to A A \mu^+ \mu^-$. $A A \to A A \pi^+ \pi^-$,
$A A \to A A \pi^+ \pi^- \pi^+ \pi^-$. This will be discussed in our future analyses. 

\vspace{1cm}

{\bf Acknowledgments}

We would like to thank Igor Pshenichnov for the correspondence and
Christoph Mayer and Joakim Nystrand for discussions.
This work was partially supported by the 
Polish grant N N202 236640 and DEC-2011/01/B/ST2/04535
as well as by the Centre for Innovation and Transfer of Natural Sciences and Engineering Knowledge in Rzesz\'ow.
A~part of the calculations within this analysis was carried out 
with the help of cloud computer system (Cracow Cloud One\footnote{cc1.ifj.edu.pl}) 
of the Institute of Nuclear Physics (PAN).



\begin{thebibliography}{100}

\bibitem{rewiev_articles}
V.M. Budnev, I.F. Ginzburg, G.V. Meledin and V.G. Serbo, 
Phys. Rep. {\bf 15} (1975) 4;
C.A. Bertulani and G. Baur, 
Phys. Rep. {\bf 163} (1988) 29; 
G. Baur, K. Hencken, D. Trautmann, S. Sadovsky and Y. Kharlov, 
Phys. Rep. {\bf 364} (2002) 359; 
C.A. Bertulani, S.R. Klein and J. Nystrand, 
Ann. Rev. Nucl. Part. Sci. {\bf 55} (2005) 271; 
A.J. Baltz, G. Baur, D.~d'Enterria et al., 
Phys. Rep. {\bf 458} (2008) 1.

\bibitem{rewiev_articles_ours}
M. K{\l}usek, A. Szczurek and W. Sch\"afer,
Phys. Lett. {\bf B674} (2009) 92;
M. K{\l}usek-Gawenda and A. Szczurek,
Phys. Rev. {\bf C82} (2010) 014904;
M. K{\l}usek-Gawenda, A. Szczurek, M.V.T. Machado and V.G. Serbo, 
Phys. Rev. {\bf C83} (2011) 024903;
S. Baranov, A. Cisek, M. K{\l}usek-Gawenda, W. Sch\"afer and A. Szczurek,
Eur. Phys. J. {\bf C73} (2013) 2335;
M. K{\l}usek-Gawenda and A. Szczurek,
Phys. Lett. {\bf B700} (2011) 322;
M. K{\l}usek-Gawenda and A. Szczurek,
Phys. Rev. {\bf C87} (2013) 054908;
%
M. K{\l}usek-Gawenda and A. Szczurek,
arXiv: 1309.2463 [nucl-th].

\bibitem{ZDC_STAR}
C. Adler, A. Denisov, E. Garcia, M. Murray, H. Stroebele and S. White,
Nucl. Instrum. Meth. {\bf A470} (2001) 488.

\bibitem{ZDC_ALICE}
K. Aamodt et al. (ALICE Collaboration), 
JINST 3 (2008) S08002.

\bibitem{Llope:1990vp} 
W.J. Llope and P. Braun Munzinger,
Phys. Rev. {\bf C41} (1990) 2644.

\bibitem{Pshenichnov:2011zz} 
I.A. Pshenichnov,
Phys. Part. Nucl. {\bf 42} (2011) 215.

\bibitem{Baur:2003ar}
  G.~Baur, K.~Hencken, A.~Aste, D.~Trautmann and S.~R.~Klein,
  Nucl.\ Phys.\ A {\bf 729} (2003) 787.

\bibitem{Plujko}
V.A. Plujko, R. Capote and O.M. Gorbachenko, 
At. Data and Nucl. Data Tables {\bf 97} (2011) 567.

\bibitem{Speth}
J. Speth and A. van der Woude, 
Rep. Progr. Phys. {\bf 44} (1988) 719.

\bibitem{GDRparameters}
V.V. Varlamov, M.E. Stepanov and V.V. Chesnokov, 
Izv. Ross. Akad. Nauk, Ser. Fiz. {\bf 67} (2003) 656.

\bibitem{Chadwick}
M.B. Chadwick, P. Oblozinsky, P.E. Hodgson and G. Reffo, 
Phys. Rev. {\bf C44} (1991) 814.

\bibitem{Vidovic}
M. Vidovi\'c, M. Greiner and G. Soff,
Phys. Rev. {\bf C48} (1993) 2011. 

\bibitem{photoabsorption_data}
J. Ahrens,
Nucl. Phys. {\bf A446} (1985) 229c.

\bibitem{gemini} 
R.J. Charity,  
Phys. Rev. {\bf C82} (2010) 014610.

\bibitem{refhaus}
W. Hauser, H. Feshbach, 
Phys. Rev. {\bf 87} (1952) 366.

\bibitem{moller}
P. M\"{o}ller, J.R. Nix, W.D. Myers, and W.J. Swiatecki, 
At. Data Nucl. Data Tables 59 (1995) 185.

\bibitem{1n:1972}
L. M. Young, Phd dissertation (1972), University of Illinois at Urbana-Champaign ("Photoneutron Cross Sections and Spectra 
from Monoenergetic Photons on Y, Pr, Pb, and Bi in~the Giant Resonance").

\bibitem{1n:1978}
R. Van de Vyver, J. Devos, H. Ferdinande, R. Carchon and E. Van Camp,
Z.Phys. {\bf A284} (1978) 91.  

\bibitem{1n:1985_1}
S.N. Belyaev, A.B. Kozin, A.A. Nechkin, V.A. Semenov and S.F. Semenko,
Yad. Fiz. {\bf 42} (1985) 1050.

\bibitem{1n:1985_2}
S.N. Belyaev, O.V. Vasilev, A.B. Kozin, A.A. Nechkin and V.A. Semenov,
Program and Theses, Proc.35th Ann. Conf. Nucl. Spectrosc. Struct. At. Nuclei, Leningrad, (1985) p.351.  

\bibitem{1n:1991}
S.N. Belyaev, A.A. Nechkin and V.A. Semenov,
Izv. Rossiiskoi Akademii Nauk, Ser.Fiz. 55 (1991) 953.

\bibitem{1n:1993}
V.V. Varlamov, N.G. Efimkin, B.S. Ishkhanov and V.V. Sapunenko,
Nuclear Constants 1 (1993) 52.

\bibitem{Livermore_1964}
R.R. Harvey, J.T. Caldwell, R.L. Bramblett and S.C. Fultz,
Phys. Rev. {\bf B136} (1964) 126.

\bibitem{Saclay_1970}
A. Veyssiere, H. Beil, R. Bergere, P. Carlos and A. Lepretre,
Nucl. Phys. {\bf A159} (1970) 561.
   
\bibitem{2n:2003}
V.V. Varlamov, N.N. Peskov, D.S. Rudenko and M.E. Stepanov,
YK (1-2) (2003) 48.

\bibitem{ALICE2012}
B. Abelev et al. (ALICE Collaboration),
Phys. Rev. Lett. {\bf 109} (2012) 252302.

\bibitem{Pshenichnov:2001qd}
  I.~A.~Pshenichnov, J.~P.~Bondorf, I.~N.~Mishustin, A.~Ventura and S.~Masetti,
  Phys.\ Rev.\ C {\bf 64} (2001) 024903.

\bibitem{Golubeva:2005nq} 
  M.~B.~Golubeva, F.~F.~Guber, T.~L.~Karavicheva, E.~V.~Karpechev, A.~B.~Kurepin, A.~I.~Maevskaya, I.~A.~Pshenichnov and A.~I.~Reshetin {\it et al.},
  Phys.\ Rev.\ C {\bf 71} (2005) 024905.

\end{thebibliography}
\end{document}